# Hedgehog Skyrmion Bubbles in Ultrathin Films with Interfacial Dzyaloshinskii-Moriya Interactions

Javier F. Pulecio[1,2], Aleš Hrabec[3,4], Katharina Zeissler[3], Ryan M. White[2], Yimei Zhu[1], and Christopher H. Marrows[3]

[1]Department of Condensed Matter Physics, Brookhaven National Laboratory, Upton, New York 11973, USA

[2]National Institute of Standards and Technology, Boulder, Colorado 80305, USA

[3]School of Physics and Astronomy, University of Leeds, Leeds LS2 9JT, United Kingdom.

[4]Present Address: Laboratoire de Physique des Solides, CNRS, Universités Paris-Sud et Paris-Saclay, 91405 Orsay Cedex, France

Collective spin ensembles known as skyrmion bubbles can exist in ultrathin magnetic film heterostructures and promise miniscule diameters and low-power all-electrical manipulation. The nucleation, identification and manipulation of skyrmions is of great interest, in part, because of their non-trivial topology. Here we ascertain the topology of 100 nm isolated hedgehog skyrmion bubbles and probe their response to external fields at room temperature in [ Pt \ Co (0.8 nm) \ Ir ] films. We find that structural disorder can stabilize skyrmion bubble states which exhibit creep-like domain wall motion. This makes the extraction of DMI from observations difficult and could explain the larger than predicted bubble sizes observed in interfacial DMI films. Intriguingly, we report on a critical size for hedgehog skyrmions, above which the bubbles lose circular symmetry and take up an irregular shape. Combining Lorentz TEM and fully three dimensional multi-layered simulations, we demonstrate the unique behavior of these topologically non-trivial spin textures.

Magnetic materials with Dzyaloshinskii-Moriya interactions (DMI), also known as antisymmetric exchange, have been predicted to form unique spin textures including skyrmions[1]. Magnetic skyrmions have a characteristic 3-dimensional spin texture with a skyrmion winding number $S$ of value 1, and so are topologically non-trivial. These topologically protected spin textures are observed as stereographic projections in thin film experiments[2] and known to exhibit distinctive properties such as a topological Hall effect[2]. They may also require orders of magnitude lower current densities to translate[3–5], unlike magnetic quasiparticles with fractional skyrmion winding numbers such as a vortex ($S$=1/2)[6–11]. While bubble domains have been explored in the past[12–14], these micron-sized domains were stabilized via magnetostatic interactions and are unworkable with current memory densities, amongst other technological challenges. On the other hand, the antisymmetric exchange interaction at the heart of stabilizing magnetic hedgehog skyrmions, may afford 5 nm bubble diameters[15] and low power manipulation[3–5], reinvigorating the possibility of magnetic bubble technologies using skyrmion bubbles[16].

Magnetic skyrmions were originally discovered at low temperatures in MnSi[17], a bulk material which lacks the crystal inversion symmetry and thus exhibits the DMI. In the thin films that are needed for any microelectronic applications, it is challenging to control the chirality of the crystalline domains so that the overall corresponding skyrmion chirality occupation is the same[18]. This problem can be eliminated in ultrathin magnetic films where the broken spatial symmetry arises from the interface between a ferromagnet and a heavy metal. The sign and magnitude of the DMI then dictates the chirality and is also believed to influence the size of the skyrmions. Interestingly, the sizes of recently observed bubble domains which are thought to have non-trivial topology, has varied over several orders of magnitude, from micron-sized objects capable of being resolved by Kerr microscopy[19] down to the nanometer scale[20] which requires high resolution magnetic techniques[15,21–24]. To determine the full topology of these bubble domains, the magnetization of the domain and the spin textures in the domain wall must be known. We will refer to bubble domains with a non-trivial skyrmion number (i.e. $S$ =1) as a skyrmion bubble henceforth, as defined by the topology. It is also possible that the topology of skyrmions could offer robustness to structural disorder in interfacial DMI films[5], though the degree of this effect has yet to be quantified.

Here we investigate the room temperature spin textures in interfacial DMI films via Lorentz transmission electron microscopy (LTEM) and provide insight regarding the magnetization of the domains as well as spin textures of the domain walls. Furthermore, we use scanning probe microscopy (SPM) and LTEM to establish the effect of local structure on the magnetic evolution under applied fields. Fully 3-dimensional, multilayered, micromagnetic simulations are implemented based on experimentally determined parameters to model our systems (see Methods for details). While the zero-field simulations show that small 25 nm diameter ground state skyrmion bubbles are stable for a narrow range of DMI values, the sizes are much smaller than anything we have observed post demagnetization. In fact, we find that applying external fields we can expand these bubbles to a critical size larger than the ground state diameters, after which they lose circular symmetry. Our results suggest the structural disorder found in our DMI heterostructures stabilizes what would otherwise be transitional states. This could account for the observations of larger metastable bubble diameters and would make the extraction of accurate DMI values from imaging alone difficult.

## RESULTS AND DISCUSSION

**Ultrathin Film DMI Heterostructures -** A schematic of the ultrathin film heterostructures used in this work is presented in Figure 1 (a) where the essential stack consisted of [ Pt \ Co \ Ir ] and was repeated 2 or 5 times to form a multilayer, henceforth referred to as a 2× or 5× sample, respectively. The DMI is a consequence of the interfacial symmetry breaking of dissimilar materials and spin-orbit coupling of the 3d Co ferromagnetic layer (FM) and the 5d Pt and Ir non-magnetic heavy metal (HM) layers. The DMI coefficients $D$ of the Pt\Co and Ir\Co interfaces are believed to have oppositely signed vector quantities[25,26] resulting in additive antisymmetric exchange and a larger effective DMI. The heterostructures were deposited on silicon nitride membranes for transmission electron microscopy (TEM) as well as on silicon\silicon dioxide wafer pieces for SQUID vibrating sample magnetometry (VSM) [27]. Figure 1 (b) shows a cross-sectional TEM image of the highly ordered heterostructure, as noted by the lattice fringes, where we measured an averaged thicknesses for the stack repetition of [ Pt 1.4 nm \ Co 0.8 nm \ Ir 0.4 nm ]. Having determined the thicknesses of the multilayers down to the atomic level, we used the magnetic volume of the FM layers to calculate the magnetic saturation moment of Co to be $M_s$ = 0.9 $\pm$ 0.2 MA m$^{-1}$ and an effective perpendicular anisotropy value of $K_{eff}$ = 0.35 $\pm$ 0.01 MJ m$^{-3}$. This level of exactness is required if micromagnetic simulations are to be used for gaining insight of the DMI from magnetic imaging, as the model is very sensitive to values of $M_s$, $K$, and $A$[5,23].

**Imaging Spin Textures using Lorentz TEM** – Lorentz TEM (LTEM) is an ultra-high resolution magnetic imaging technique and sensitive to the magnetization in minute volumes. In Fresnel mode, LTEM is sensitive to the effect of a magnetic field on the electron wave front. This effect is often simplified to a **v** × **B** Lorentz force product which we will use for explanation purposes below, but we note for proper treatment one must consider the electron motion as a complex wave function whose phase is affected by the magnetic field, as Fresnel imaging is strictly carried out under defocused imaging conditions[28]. Supplementary Figure 1 shows how magnetic contrast is formed via LTEM electron beam interactions in an over-focused condition and was recently reported by Benitez et al.[29] in Pt | Co | AlO$_x$ systems with homochiral walls. For out-of-plane (OOP) magnetized domains, the electron path is not affected by the magnetization at normal incidence (i.e. zero tilt), and the standard (gray) amplitude contrast lacks any magnetic features. At the Bloch wall though, where the magnetization rotates in-plane, the electron path is altered creating an absence of intensity (black) on the detector directly under the wall and increases the intensity to the right of the wall (white) due to the superposition of the unaffected electrons passing through the domain. Under the same zero tilt configurations, a Néel wall does *not* create similar contrast because the electrons are deflected along the length of the wall. If the sample is then tilted, the Néel wall still does not contribute to the resultant contrast but a portion of the OOP magnetization from domains creates a single dark spot directly under the wall.

Figure 1 (b) shows an experimental LTEM image of a skyrmion bubble found in our films, where the yellow 300 nm scale bar shows the diameter of the bubble. The red and blue false color was overlaid to highlight the respective up and down OOP domains. The white and black LTEM contrast shows the position of the Néel wall which was only revealed when the sample was tilted. By determining the direction of magnetization of the domains as well as the type of wall present in these films, we can establish the full topology to be that of a hedgehog skyrmion. We note though that LTEM imaging cannot determine the handedness of the skyrmion, which is set by the sign of $D$. The handedness can be inferred through the dependence on the stacking order of the interfacial DMI. For our films, the left-handed chirality shown in the schematic below the skyrmion in Figure 1 (b) is based on a negative $D$ which arises from the Pt \ Co \ Ir stacking order. As discussed earlier, the chirality for the hedgehog skyrmions in interfacial DMI films can be engineered by the symmetry breaking at the interface.

**Magnetic Topology of Ultrathin DMI Films –** Before proceeding, it is valuable to model the effects of the DMI and identify the distinctive magnetic phases. In Figure 2 we have used the experimentally determined magnetic parameters and multilayer stack thicknesses shown in Figure 1, while varying $D$ in the OOMMF package[30] with the DMI extension[31]. The simulations were initialized with a hedgehog skyrmion bubble of diameter $d$ = 20 nm and the figure shows the spin textures after a relaxation time of 15 nanoseconds in a zero field environment. The contrast scheme used was selected to accentuate the Néel domain walls. For $|D| \leq 1.5$ mJ m$^{-2}$, as depicted in Figure 2 (a), the hedgehog skyrmion bubble collapses and the system settles into a single domain ground state. A DMI coefficient of $|D|$ = 2.0 mJ m$^{-2}$ yields a hedgehog skyrmion ground state with a diameter of 25 nm. Figure 2 (b) shows an unstable state with a skyrmion bubble for $|D|$ = 2.5 mJ m$^{-2}$ and the corresponding DMI ($D$), Heisenberg exchange ($A$), anisotropy ($K$), and dipolar ($D_m$) energies plotted in blue in Figure 2(e-h), respectively. In this regime, the bubble appears to be stable with very little change in the magnetic energies with a stopping criteria of d$M$/d$t$ = 1.0, but expands beyond the simulated geometry of 2 μm for a d$M$/d$t$ = 0.1 after 20 ns. Therefore, caution should be used when determining the ground states to allow the simulation to evolve under the appropriate stopping conditions.

The seeded bubble breaks down for larger values of $|D|$ > 3.0 mJ m$^{-2}$ into labyrinth domain patterns as shown in Figure 2 (b-c), with the magnetic energies plotted in green and red for $|D|$ = 3.5 mJ m$^{-2}$ and $|D|$ = 4.5 mJ m$^{-2}$, respectively. Figure 2 (c) also outlines the node-branch configurations found in the labyrinth domain patterns. As the magnitude of $D$ is increased, the resulting width of the branches are reduced. These spin textures clearly show the effect of $D$ and results from the energy minimization of all the energies calculated in the model. Figure 2 (e-h) plots the total energy for $D$, $A$, $K$, and $D_m$ as the system approaches a global energy minimum over time. As $A$, $K$, and $D$ approach competing energetic magnitudes, their interplay is what leads to the distinct phases.

**Local Structure and Ground states** – Under ideal circumstances, the demagnetization process used here (see Methods) should allow for the system to reach a global energy minimum. Indeed, this has been used to help understand inherent properties of the DMI heterostructures[23]. In the micromagnetic simulations presented in Figure 2, the model assumes no structural disorder and only the inherent magnetic energies determine the global ground states. In the physical system, however, structural imperfections exist and play a role in the magnetic states observed post demagnetization. Figure 3 (a-c) shows a 5× heterostructure surveyed using scanning probe microscopy. Figure 3 (a) shows the topographic image of the film's surface acquired in the non-contact mode of atomic force microscopy and was the initial scan (s=0) in a series of scans. The blue and red arrows are used to point out several structural defects found on the surface of the film stack. Figure 3 (b) then uses a lift-mode phase scan to probe long-range magnetostatic interactions with the magnetic tip (magnetic force microscopy). The brighter areas correspond to magnetic bubbles that appear in the film and the white arrow shows the slow scan direction. The red arrow shows a magnetic bubble that was stable partly through the scan, until it was sufficiently perturbed by the tip's magnetic field.

Figure 3 (c) shows a subsequent image where the offset distance is reduced so as to include both magnetostatic and van der Waals forces phase information. As outlined by the blue arrows, there exists a spatial relation between the bubbles and structural defects. The red arrow shows a structural defect in (a & b) that served to anchor the bubble that was disturbed by the scan in (b) and no longer present in the subsequent scan (c). This suggests the structural disorder in the film can serve as pinning/nucleation site for the magnetic bubbles. While it is difficult to correlate the structural defect size to the magnetic bubble size, the film does exhibit magnetic bubbles on the order of 100 nm which is smaller than traditional bubbles stabilized via magnetostatics in materials such as orthoferrites[12–14] and evidence of isolated bubbles in this material system.

**The Effect of OOP External Fields on Interfacial DMI films –** In conventional garnet bubble systems with perpendicular magnetic anisotropy (PMA) and negligible DMI, the process for bubble nucleation is well understood. After demagnetization, OOP stripe domains with Bloch type walls are preferred. These systems can be classified by the transitions that occur as a function of the applied OOP field strength. As a field is applied, at a certain value $B_l$ the stripe domains give way to a bubble lattice which may lead to complex domain wall structure[32]. The field strength can be further increased to the critical bubble collapse field $B_c$. If the field is reduced prior to the bubble collapse field another critical field exists where the bubbles begin to 'strip out' $B_{so}$ back into stripe domains. In the following we discuss how the systems studied in this work behave in external fields with the addition of DMI that stabilizes Néel walls forming homochiral walls and hedgehog skyrmions.

Post-demagnetization, the DMI heterostructure with two stack repetitions favored large domains and also exhibited a few isolated hedgehog skyrmion bubbles, as shown in Supplementary Figure 2. In Figure 4 (a-c) we applied OOP fields to shrink the larger domains to see if the system would promote the formation of skyrmion bubbles, similar to conventional bubble systems. As can be seen in Figure 4 (a) after applying a field of $B_z$ = 15.0 mT the remanent domains collapse such that only the Néel walls are visible and adjacent to one another with the node-branch type configuration shown in Figure 2 (c). The homochiral Néel walls provide an extra energy barrier preventing the annihilation of the walls[29]. In Figure 4 (b) a larger field of $B_z$ = 15.5 mT was applied, where a percentage of the Néel walls have mutually annihilated and a skyrmion bubble was created (red arrow). We have seen very few bubbles nucleate in this fashion and suspect that local disorder plays a role in this process as previously discussion for Figure 3.

In Figure 4 (c) the field was increased slightly to $B_z$ = 15.9 mT which was near saturation, where more of the node-branch structures have annihilated. Notably, unlike conventional bubble materials, the onset of a bubble lattice is not realized. Most of the node-branch spin textures with adjacent homochiral Néel walls prefer annihilation rather than bubble nucleation. Another difference worth noting is that the hedgehog skyrmion bubbles found in these films do not 'strip out' into stripe domains once the field is removed (not shown here), and unlike bubbles found in traditional materials, are stable in zero field once nucleated.

We then proceeded to reverse the field direction such that $B_z$ is now parallel to the core of the skyrmion bubble in Figure 4 (d-f), and so will cause the bubble to inflate. Figure 4 (d) shows that the stripe-like domains expand more readily than the skyrmion bubble for $B_z$ = -9.7 mT. This could be the result of a combinations of factors including the energetic cost of expanding a bubble domain with circular symmetry versus a stripe domain as well as variations in the local structure causing pinning. In Figure 4 (e) the field was increased to $B_z$ = -11.5 mT and now both the stripe-like domains as well as the skyrmion bubble have increased in size. Importantly, this clearly demonstrates an independence of the skyrmion bubble from neighboring domains unlike a node connected by the stripe domain branches. Furthermore, it is clear that the annihilation of Néel walls is a non-reversible process when these images are compared to Figure 4 (a). The field is then increased to $B_z$ = -15.9 mT where the bubble loses its circular symmetry. This appears to be similar to an inverse 'strip out' process, where instead of reducing the field gives rise to a non-bubble state, such as a stripe domain in conventional bubble films, we are actually increasing the field parallel to the bubble domain in such a way that the bubble breaks down into a lower order of symmetry.

**Hedgehog Skyrmion Bubble Breakdown** – Using the process outline in the Methods section, we demagnetized several films and directly imaged the remanent states using LTEM. We found magnetic configurations very similar to those reported by Moreau-Luchaire *et al.*[23] for a high number of repetitions (10×) as shown in Supplementary Figure 2 (a). The remanent magnetic arrangement for those films consisted of labyrinth domains of approximately 100 nm in width and so exhibited a high density of domain walls. While we found the field evolution to be complex, high resolution LTEM imaging with sub 10 nm resolution[33] allowed us to see the domains contract similar to what is shown in Figure 4 (a-c). We would add that node-branch magnetic configurations of the labyrinth patterns discussed in Figure 2 are prevalent in the 10× films. If the domains are contracted such that the branch widths are near or below the resolution of the imaging technique, the nodes could appear as isolated bubbles for techniques that are primarily sensitive to the OOP magnetization. In contrast, observations of the 2× films in Supplementary Figure 2 (b) show more expansive domains post demagnetization, as well as clearly isolated skyrmion bubbles (white arrows). In fact, we have shown that the energy of these systems can be tuned such that the topologically non-trivial skyrmion bubble phase (2× films) is preferred over the cycloidal state (10× films)[27].

Figure 5 (a) shows an isolated skyrmion bubble in a 2× DMI film after demagnetization, where the inherent dipolar energy leads to favoring the bubble phase as a ground state[27]. The black and white LTEM wall contrast only appears when the sample is tilted, indicative of Néel domain walls. By identifying the wall type and OOP magnetic domains, we can determine the bubble's magnetic topology as a full hedgehog skyrmion as discussed for Figure 1 (b). The yellow scalebar in Figure 5 (a) is 115 nm in length and is the diameter of the skyrmion bubble. In general, this was the minimum skyrmion diameter we could find in our films post-demagnetization. In (a-c) we perturbed the bubble by providing 5 sec field pulses ($n_p$ = the number of pulses applied) at 11.5 mT and subsequently observed the remanent states. The observed diameter of the skyrmion was increased in (b & c) while maintaining the bubble's circular symmetry.

The field magnitude was then increased to 12.8 mT, and the same pulse protocol followed, with the results shown in (d-f). The bubble structure loses its circular symmetry and ultimately breaks into irregular domain shapes as shown in Figure 5 (e) and (f). The critical diameter at which the skyrmion bubble bursts into an asymmetric domain was $d_c$ = 550 nm, as determined by the last observed bubble size prior to breakdown. A similar effect is well known in soap bubbles, which lose spherical symmetry and become irregular above a critical size. This point is determined by the size above which surface tension is no longer the predominant force determining the bubble shape. In the case of our magnetic bubble the domain wall motion appears to fall into a thermally activated creep/depinning regime[34], where the local disorder creates local energy minima allowing for the observation of these transitional states. These pinning forces now play an important role in determining the bubble shape. It may seem intuitive to correlate the bubble breakdown size from the experiment to the micromagnetic simulations, however, the structural disorder and the Zeeman energy necessary to evolve the experimental system alters the energy landscape, and therefore, makes the directly correlation to simulations without defects unworkable. Therefore, including local disorder into the models and understanding its effects on domain wall motion in interfacial DMI films should prove to be relevant.

## Conclusion

We have observed and modeled the chiral spin textures found in interfacial DMI ultrathin film heterostructures of [ Pt \ Co (0.8 nm) \ Ir ] at room temperature. LTEM imaging was used to map the magnetic topology of hedgehog skyrmions and labyrinth domains patterns which we inferred to have left-handed homochiral walls based on the stacking order. We demonstrated via SPM and LTEM that the domain walls exhibit sensitivity to the structural disorder found in the films with creep-like motion, and that pinning can sustain larger-then-equilibrium sub-micron skyrmion bubble states. By applying OOP external magnetic fields parallel to the bubble domain core, we grew the hedgehog skyrmions bubbles to a critical size of 550 nm, after which the circular symmetry broke down into irregularly-shaped domain patterns. We conclude that the structural variation found in our interfacial DMI films can cause larger than expected skyrmion bubble sizes, making correlations from observed remnant states to micromagnetic simulations informative, but insufficient for the extraction of the Dzyaloshinskii-Moriya interaction energy. Moreover, the process to nucleate skyrmion bubbles seems to be influenced by local structure and affects the Néel domain wall motion that are stabilized by the interfacial DMI. Having directly determined the topology of the hedgehog skyrmion bubbles, we have demonstrated that these skyrmion bubbles can be isolated in thin films. Furthermore, the topologically non-trivial skyrmion bubbles found in our films can be stabilized in zero field as a ground state and behave differently than bubbles found in traditional PMA materials with negligible DMI.

## Methods

### Sample Preparation and Characterization

The heterostructures of Ta(3.2 nm)\Pt(3 nm)\[Co(0.8 nm)\Ir(0.4 nm)\Pt(1.4)]×N were deposited by dc magnetron sputtering at base pressure $10^{-8}$ Torr on silicon nitride membranes for transmission electron microscopy (TEM) as well as on thermally oxidized silicon wafer pieces for conventional MOKE and SQUID magnetometry. The anisotropy constant $K$ and magnetization $M_s$ were measured using a superconducting quantum interference device vibrating sample magnetometer (SQUID-VSM).

## Demagnetization

The demagnetization process used for the LTEM investigations consisted of applying an alternating OOP field, normal to the sample surface, at a magnitude of 1 T down to zero. A saturating 2 T field parallel to the sample surface along the hard magnetization direction was also used for the SPM experiments.

## Lorentz Imaging

LTEM was performed at 200 kV using the JEOL 2100F-LM, a dedicated Lorentz microscope with a weakly excited objective lens (< 4 Oe magnetic field on the sample). The skyrmion bubble diameters were defined by measuring the 180° peak-to-peak domain wall intensities.

## Micromagnetic Simulations

Micromagnetic simulations were carried assuming ideal circumstances neglecting effects of structural variations and temperature using the NIST provided Object-Oriented Micromagnetic Framework (OOMMF) Oxsii solver[30] with the DMI extension[31]. A fully 3-dimensional, multilayered, micromagnetic model was implemented to capture the individual layers and interfaces as they were deposited in the experiment. We found this fully 3D multilayered numerical approach to provide distinct solutions from works employing effective single layer models[23,24]. The micromagnetic simulations parameters of $M_s$ = 0.985 MA m$^{-1}$, $A$ = 15.0 pJ m$^{-1}$, $K_{eff}$ = 0.35 MJ m$^{-3}$, $D$ = -1.5 to -4.5 mJ m$^{-2}$, and a cell size of 2 nm × 2 nm × 0.8 nm were used under zero applied field for a film stack of [ FM(0.8 nm) \ HM(1.6 nm) ]$_{×2}$.

**Acknowledgements**
We would like to acknowledge support by the US Department of Energy, Basic Energy Sciences, under Contract No. DE-AC02-98CH10886 and the EPSRC through grant numbers EP/L00285X/1 and EP/I011668/1. This work also received support from the European Union's Horizon 2020 research and innovation program under grant agreement No 665095 (MAGicSky).

**Author contributions**
J.F.P. and C.H.M. conceived the study. A.H. grew the additive multilayer films. K.Z. and A.H. performed the magnetic characterization of the films. J.F.P. performed the LTEM imaging and interpreted the results with Y.Z. and C.H.M. The SPM data was acquired and analyzed by J.F.P. J.F.P performed the micromagnetic simulations. R.W. and J.F.P performed and analyzed the HR-TEM of the cross-section. All authors contributed to the interpretation of the results and writing of the manuscript.

**Additional information**
Supplementary information is available in the online version of the paper. Reprints and permissions information is available online at www.nature.com/reprints. Correspondence and requests for materials should be addressed to J.P.

**Competing financial interests**
The authors declare no competing financial interests.

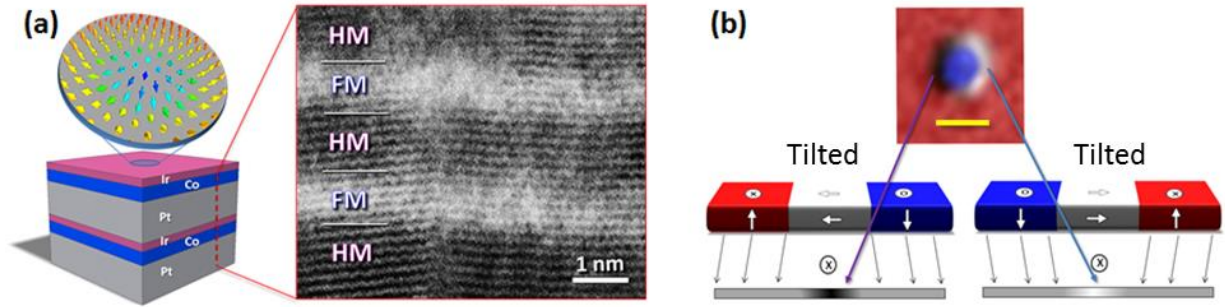

Figure 1 – Ultrathin film DMI heterostructures. (a) A schematic of the ultrathin film heterostructure which is designed to contain hedgehog skyrmion bubbles shown in the magnetic vector plot above. A high resolution STEM bright-field cross sectional image of the stack shows the ferromagnetic (FM) and heavy (HM) metal lattice fringes. The thicknesses of the metals were measured and used to calculate the magnetic properties of the heterostructure via conventional magnetometry. (b) Shows an experimental Lorentz TEM image of a skyrmion where the red and blue color overlays are used to relay the magnetization direction of the domains and the yellow scale bar is 300 nm. The black and white LTEM contrast represents the position of the Néel domain walls.

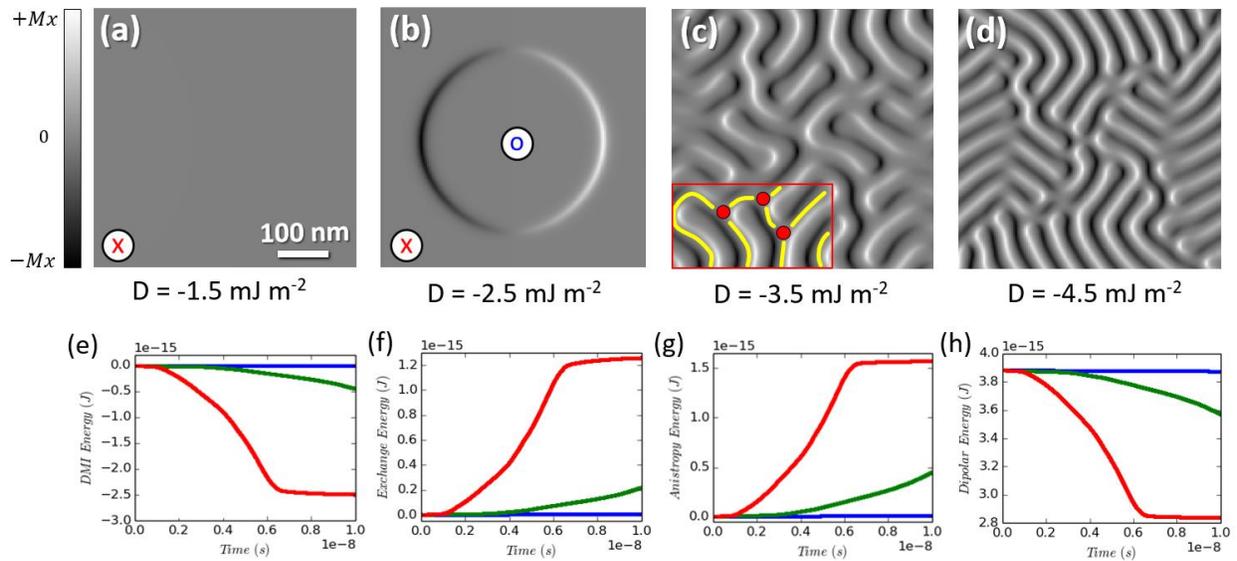

Figure 2 – Simulated effect of DMI in ultrathin film systems. (a) Single domain (red x - down), (b) hedgehog skyrmion bubble (blue o - up), and (c-d) labyrinth domain phases at t = 15 ns for different values of $D$. The overlay in the lower of left corner of (c) outlines the node (red circle) branch (yellow lines) configurations of the labyrinth domains. The total energies integrated over the magnetic volume for the $D$ = -2.5 (blue), -3.5 (green), and -4.5 (red) mJ m$^{-2}$ simulations are shown in (e-h), where each simulation was seeded with a skyrmion bubble 20 nm in diameter.

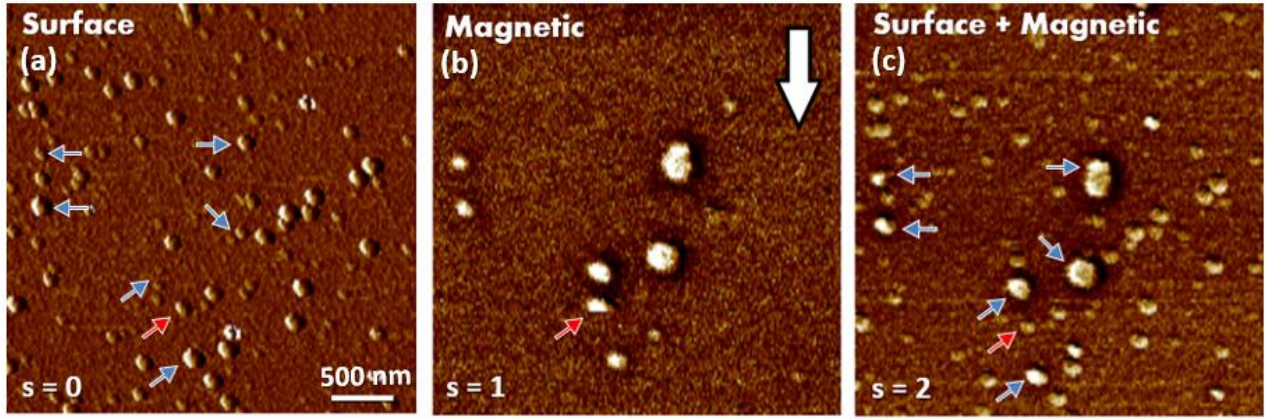

Figure 3 – Effects of physical structure on spin textures. (a-c) SPM images of a 5× DMI heterostructure over sequential scans ($s$ = 0, 1, 2). (a) Shows the top most surface of the heterostructure with several structural defects outlined by the blue arrows. (b) Shows the magnetic bubbles present in the film. The red arrow outlines a bubble that was perturbed by the magnetic tip along the slow scan direction (white arrow). (c) Combined mapping of the surface morphology and magnetic structure showing the correlation between physical structure and skyrmion bubbles.

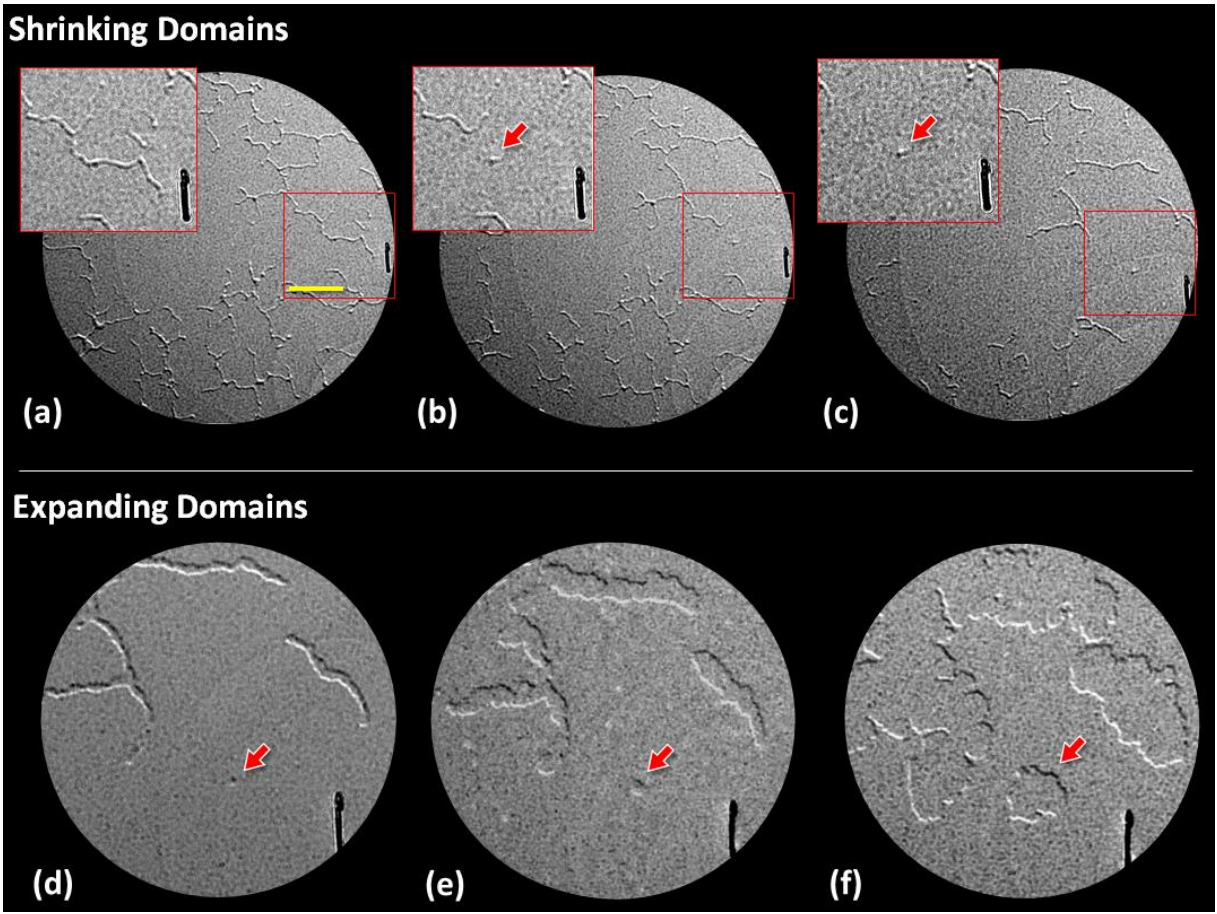

Figure 4 – Manipulating domains using external fields. (a) An out-of-plane field of $B_z$ = 15.0 mT was applied to shrink the domains until the homochiral Néel walls were adjacent. This is evident in the expanded region outlined by red overlays of the Lorentz TEM images (20 $\mu$m yellow scale bar) which clearly shows several node-branch spin configurations. (b) The field was increased to $B_z$ = 15.5 mT and was enough to annihilate some of the Néel walls. The red arrow shows a unique area where a hedgehog skyrmion bubble broke off the striped domain. (c) The field was further increase to $B_z$ = 15.9 mT where the many of the Néel walls were annihilated but the skyrmion remains just prior to full saturation. (d) A field in the opposite direction of $B_z$ = -9.7 mT and parallel to the skyrmion core was applied. The striped domains expand while the bubble appears unchanged. (e) The field is increased to $B_z$ = -11.5 and the domains expand. (f) The field is increased to $B_z$ = -15.9 and the bubble breaks down.

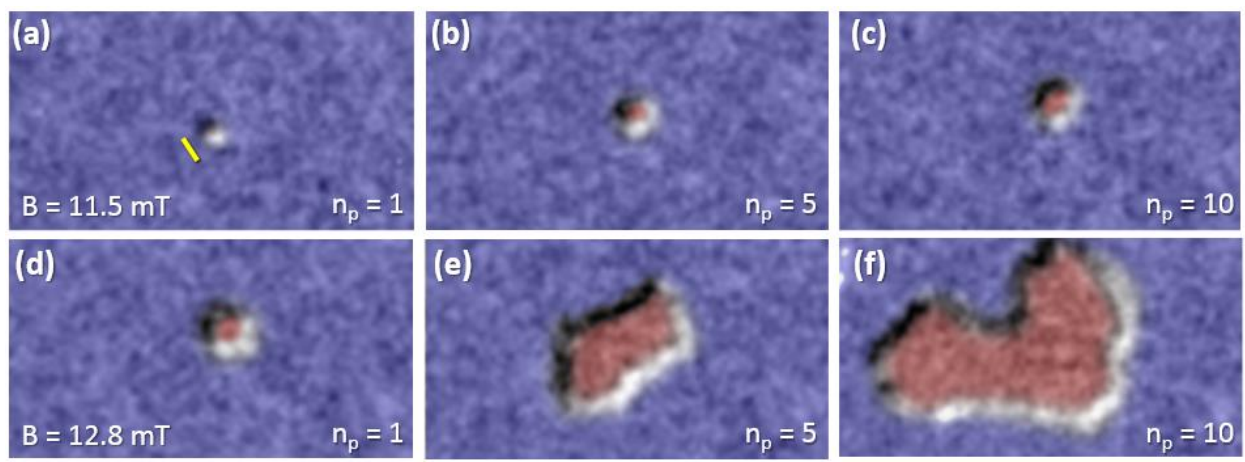

Figure 5 – Breakdown of a hedgehog skyrmion bubble. (a-f) LTEM images for a 2× heterostructure where the remnant spin textures are observed following the application of out-of-plane field pulses. The color overlay was added for clarity of the out-of-plane domains while the white and black LTEM contrast depicting the position of the Néel wall is unchanged. The field pulses ($n_p$ = number of pulses) were applied for 5 seconds with out-of-plane field magnitudes of 11.5 mT (a-c) and 12.8 mT (d-f). After a critical size of approximately 550 nm the skyrmion's circular symmetry breaks down and the domain expands in an irregular fashion.

# Supplementary Information

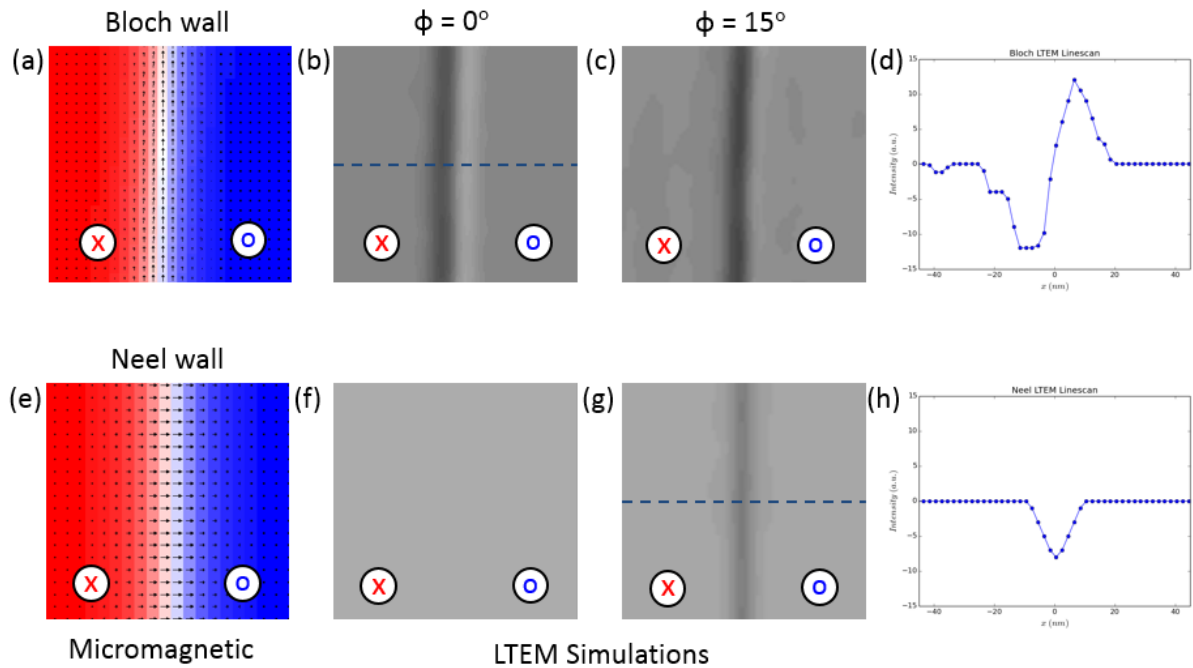

S1 – LTEM Contrast Simulations. (a) A micromagnetic simulation of a Bloch wall transition. (b-c) Zero tilt (ɸ = 0˚) and tilted (ɸ = 15˚) over-focused LTEM simulations of the micromagnetic part containing a Bloch wall. The essential LTEM contrast results from the beam interaction with the domain wall's in-plane magnetization for both tilt angles. (d) A line scan of the contrast signature for a Bloch wall transition. The in-plane moments at the domain wall form the minimum and maximum features as a result of the Bloch wall's beam interaction. (e) A micromagnetic simulation of a Néel wall transition. (f-g) Zero tilt (ɸ = 0˚) and tilted (ɸ = 15˚) over-focused LTEM simulations of the micromagnetic part containing a Néel wall. The in-plane magnetic moments shift the beam along the length of the Néel wall which lacks any differentiable contrast. Unlike the Bloch wall, when tilted the LTEM contrast originates strictly from the tilted in-plane component of the out-of-plane domains. (h) A line scan of the contrast signature for a Néel wall transition. The beam interaction with the domains themselves, and not the domain wall, creates a distinct single minimum.

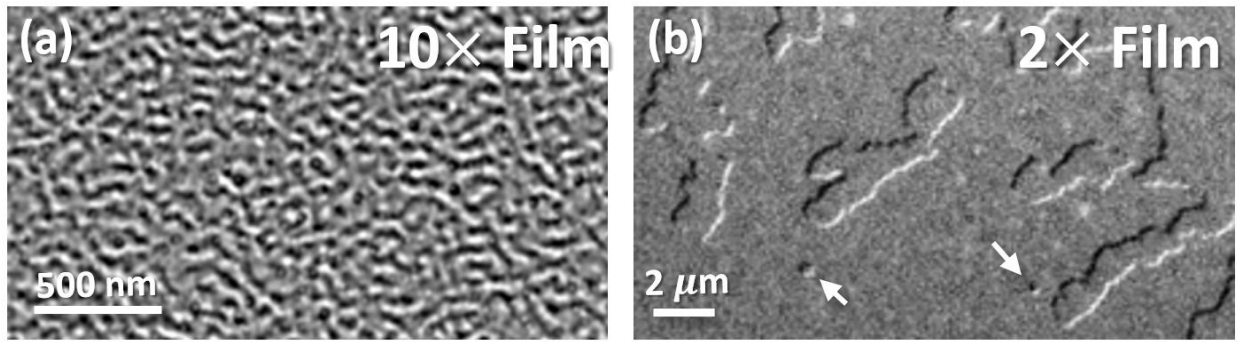

S2 – Isolation of skyrmion bubbles. (a) The 10× film shows the much more complicated and dense domain wall pattern (black and white lines), where the labyrinth domains are on the order of 100 nm in width after demagnetization.. (b) The white arrows in the 2× film point out clearly isolated sub-micron skyrmions surrounded by larger micron-sized domains that were present post demagnetization.

13